\documentstyle[twoside,fleqn,epsfig,espcrc2]{article}

\newcommand{\AmS}{{\protect\the\textfont2
  A\kern-.1667em\lower.5ex\hbox{M}\kern-.125emS}}

\title{Topology in full QCD with 2 colours at finite 
temperature and density}
\author{  B.~All\'es\thanks{Speaker 
           at the conference.}\address{Dipartimento 
           di Fisica, Universit\`a di Milano--Bicocca and INFN
           Sezione di Milano, 20126--Milano, Italy}, 
          M.~D'Elia\address{Dipartimento di Fisica, 
           Universit\`a di Pisa and INFN Sezione di Pisa, 
           56127-Pisa, Italy}, 
          M.~P.~Lombardo\address{INFN Sezione di Padova,
           e Gruppo Collegato di Trento, Italy}, 
          M.~Pepe\address{Laboratoire de Physique Th\'eorique, 
                          Universit\'e de Paris XI,
                          B\^atiment 210, 91405 Orsay Cedex, France}
}

\begin{document}

\begin{abstract}
We present preliminary results about topology and several other
quantities at non-zero baryon density and finite temperature 
in full QCD with $N_c=2$.
\end{abstract}

\maketitle

\section{INTRODUCTION}

The behaviour of the QCD vacuum at finite density can be tested in the
heavy ion colliders. Therefore it is interesting to have a 
set of predictions from QCD and the lattice is a good means to obtain them
in the non--perturbative regime.

We have analysed the model with two colours in order to make use of
standard simulation algorithms. We expect that this simplification 
may leave at least some of the essential features of 
the problem. 

We give a progress report of a simulation performed on a $14^3\times
6$ lattice by using 8 flavours of staggered fermions at masses
$am=0.05$ and $am=0.07$ at several values of $\beta\equiv 4/g_0^2$ where
$g_0$ is the lattice bare coupling. The Monte Carlo 
algorithm is the standard hybrid molecular dynamics~\cite{duane}. 
The action is~\cite{hasenfratz}
\begin{eqnarray}\label{action}
 S &=& \frac{1}{2} \sum_n \eta_4(n) \Big\{
                  {\rm e}^{a\mu} \overline{\psi}(n)
                  U_4(n)\psi(n+\widehat{4}) \nonumber \\
   &&- {\rm e}^{-a\mu} \overline{\psi}(n+\widehat{4})
                  U_4(n+\widehat{4})\psi(n) \Big\} \nonumber \\
   &&+ \hbox{spatial terms,}
\end{eqnarray}
where the ``spatial terms'' are those independent of the 
chemical potential $\mu$ and $\eta_4(n)$ are the staggered phases.

Our results are preliminary because {\it i)} we have not yet
calculated all the renormalization constants necessary for the
evaluation of the physical topological susceptibility 
$\chi$~\cite{alles1} and {\it ii)} we have
not calculated the physical units by determining the lattice spacing~$a$.

\section{PARTICLE DENSITY}

In Fig.~1 we show the particle density $\rho$ as a function of the
chemical potential $\mu$ for masses $am=0.05$ and $am=0.07$. 
In both cases we see a cubic dependence for large values of $\mu$
that is in agreement with the analytical expectation~\cite{muller},
\begin{equation}
\label{density}
\rho = {8\over 3}\; \left[\mu \; T^2 + {\mu^3\over \pi^2}\right]
       \; .
\end{equation}
for $\mu\gg T$ where $T$ is the temperature.

\begin{figure}[t]
\centerline{\epsfig{file=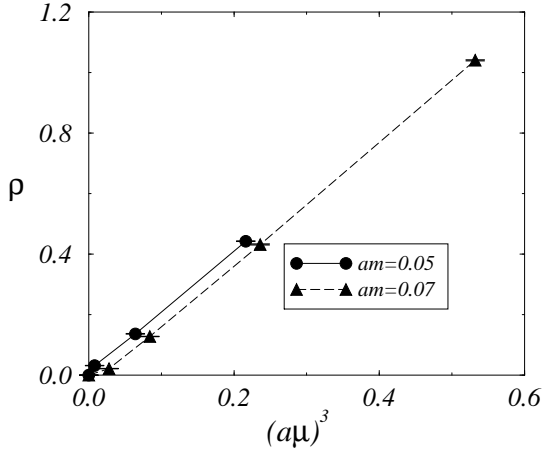,angle=-90,width=0.45\textwidth}}
\caption{Baryon density $\rho$ as a function of the chemical potential
at $\beta=1.5$.}
\end{figure}

\section{TOPOLOGICAL SUSCEPTIBILITY}

The topological susceptibility $\chi$ is an important parameter of the QCD
vacuum. In the continuum and for the $SU(2)$ gauge group it is defined as
\begin{equation}
 \chi \equiv \int d^4x \;\; \partial_\mu 
   \langle 0 | {\rm T}\left\{K_\mu(x) Q(0)\right\}| 0 \rangle \; ,
\label{eq:chi}
\end{equation}
where $K_\mu(x)$ is the Chern current
\begin{equation}
 K_\mu(x) = {g^2 \over 16 \pi^2} \epsilon_{\mu\nu\rho\sigma} A_\nu^a 
              \partial_\rho A^a_\sigma \; ,
\end{equation}
$g$ is the QCD coupling constant and $Q(x)=\partial_\mu K_\mu(x)$ is
the topological charge density,
\begin{equation}
 Q(x) = {g^2\over 64\pi^2} \; \epsilon^{\mu\nu\rho\sigma} \;
        F^a_{\mu\nu}(x) \; F^a_{\rho\sigma}(x)  \; .
\end{equation}
Eq.~(\ref{eq:chi}) uniquely defines the prescription for the singularity 
of the time ordered product when\break $x\rightarrow 0$~\cite{witten}.

On the lattice, the evaluation of $\chi$ does not follow the above
prescription. As a consequence we have to subtract the singularity
which shows up as an additive renormalization. This subtraction is
performed following the method of Ref.~\cite{alles1}.

It is well--known that $\chi$ undergoes a sharp drop at the
deconfinement temperature, both for three colours~\cite{alles1} and
for two colours~\cite{alles2}. We have measured this quantity as a
function of the chemical potential in order to study its behaviour at
finite density as well as finite temperature. In Fig.~2 we show
\begin{figure}[htbp]
\centerline{\epsfig{file=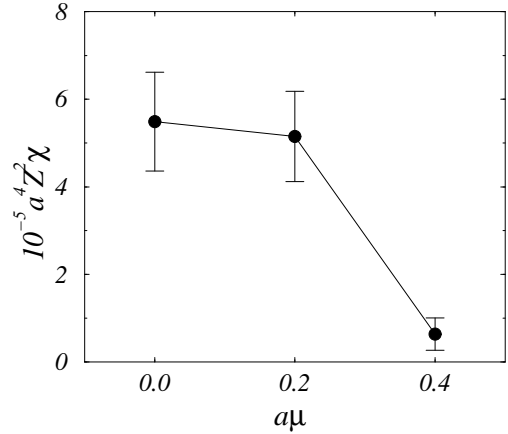,angle=-90,width=0.4\textwidth}}
\caption{Topological susceptibility as a function of the chemical
potential at a finite temperature, $\beta=1.5$ and $am=0.07$.}
\end{figure}
that the behaviour is qualitatively analogous to the
above--described one at finite temperature and zero density. 
The quantity $Z$ in
the vertical axis is the finite renormalization which relates the
topological charge in the continuum and on the
lattice~\cite{campostrini}. We are presently running more statistics to
sweep the sector beyond the critical density and to improve the error bars.

\section{THE POLYAKOV LOOP}

To investigate the deconfinement properties of the theory across the phase
transition at finite chemical potential,
we have measured the Polyakov loop $\Pi$. Our results, shown in Fig.~3,
indicate that the theory becomes deconfining beyond a critical density and
suggest that the transition occurs at $a\mu\sim 0.4$ for fixed 
temperature. A similar trend can be observed by varying the 
temperature at zero chemical potential, Fig.~4, where we have
traded the temperature $T$ for the lattice coupling $\beta$.
\begin{figure}[t]
\centerline{\epsfig{file=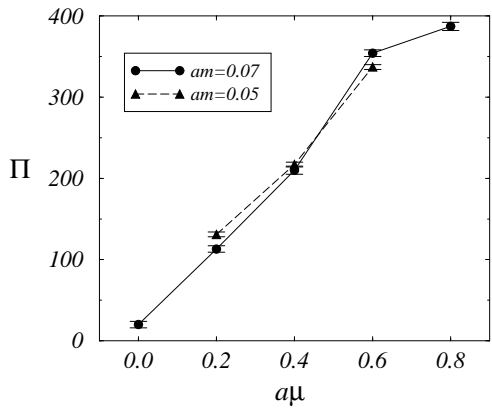,angle=-90,width=0.4\textwidth}}
\caption{Polyakov loop $\Pi$ as a function of the chemical potential
for $\beta=1.5$.}
\end{figure}
\begin{figure}[t]
\centerline{\epsfig{file=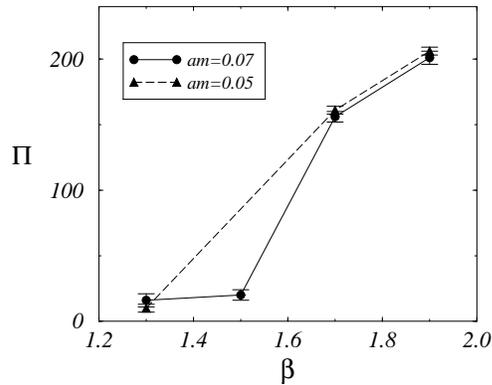,angle=-90,width=0.4\textwidth}}
\caption{Polyakov loop $\Pi$ as a function of the temperature (written as 
$\beta$) for zero chemical potential.}
\end{figure}

\section{CHIRAL TRANSITION}

The chiral symmetry is known to be restored at the critical
temperature. This result calls for a study of the behaviour of this
symmetry across the critical density. In Fig.~5 we show the chiral
condensate $\langle\overline{\psi}\psi\rangle$ as a function of the
chemical potential. We see that there is a transition approximately at
the same value of $\mu$ at which the deconfinement took place.
\begin{figure}[t]
\centerline{\epsfig{file=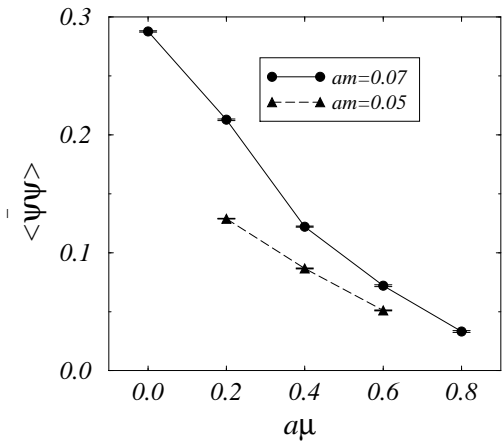,angle=-90,width=0.4\textwidth}}
\caption{Chiral condensate as a function of the chemical potential at
$\beta=1.5$.}
\end{figure}
An analogous figure for $\langle\overline{\psi}\psi\rangle$ is
obtained at zero chemical potential as a function of $T$.

\section{PARTICLE SUSCEPTIBILITIES}

The particle susceptibility $\chi_{\rm particle}$ is defined as
\begin{equation}
\chi_{\rm particle}\equiv \sum_t G(t) \; ,
\end{equation}
where $G(t)$ is the propagator of such a particle at time $t$.
If for example we consider a pion and a single
pole dominates the propagator,
then $\chi_\pi\sim Z_\pi/m_\pi^2$ where $Z_\pi$ is the pion field
renormalization constant. Moreover, if $Z_{\rm particle}$ is 
insensitive to the quantum properties of the particle, 
then $\chi_{\rm particle}$ can provide useful information about 
spectroscopic level ordering~\cite{hands}. 
Actually we shall make use of this
result to compare the masses before and after the transition and we
shall not need to be careful about the definition of $G(t=0)$.

In Fig.~6 the susceptibilities for pions and deltas are shown. 
After the transition the chirality does not matter
and the masses of parity opposite particles coincide. This is another
way to see that the chiral symmetry is restored. We see again that this
restoration occurs at $a\mu\sim0.4$ for $\beta=1.5$. 
\begin{figure}[thb]
\centerline{\epsfig{file=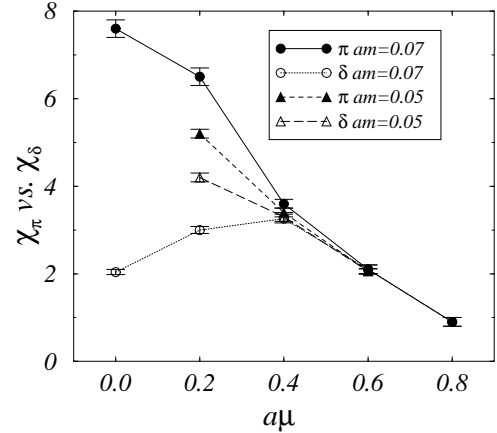,angle=-90,width=0.4\textwidth}}
\caption{Particle susceptibility as a function of the chemical potential at
$\beta=1.5$.}
\end{figure}
In Fig.~7 we see the analogous plot displaying the same phenomenon for
zero density.
\begin{figure}[htbp]
\centerline{\epsfig{file=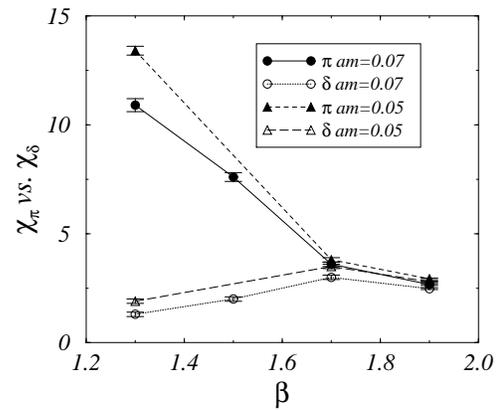,angle=-90,width=0.4\textwidth}}
\caption{Particle susceptibility as a function of $\beta$ for $\mu=0$.}
\end{figure}

\section{CONCLUDING REMARKS}

We have presented several plots obtained from a numerical simulation 
of QCD with gauge group $SU(2)$ at finite temperature and density. Although 
our results are still preliminary, we can draw a qualitative picture of the
phase diagram of the theory in the temperature--chemical potential plane
(we actually substitute the temperature for the $\beta$ value, having
fixed the temporal size of the lattice).
There is a region where confinement, chiral symmetry breaking and a 
non--zero value for the topological susceptibility occurs.
This region is bounded at $\mu=0$ by $\beta\sim 1.6$ and at $T=0$ by
$a\mu\sim 0.4$~\cite{hands}. We have checked this picture from
several measurements, all of them giving figures in agreement; for instance
the position at $\beta=1.5$ where the 
Polyakov loop raises from zero, the chiral condensate
and the topological susceptibility vanish and the particle 
susceptibilities coalesce is always $a\mu\sim 0.4$.

\end{document}